\documentclass[12pt,letterpaper]{article}
\usepackage[latin1]{inputenc}
\usepackage{amsmath}
\usepackage{amsfonts}
\usepackage{amssymb}
\usepackage{graphicx,color}

\newcommand{\al}{\alpha}
\newcommand{\ga}{\gamma}
\newcommand{\de}{\delta}
\newcommand{\ep}{\epsilon}

\newcommand{\be}{\begin{equation}}
\newcommand{\ee}{\end{equation}}
\newcommand{\bea}{\begin{eqnarray}}
\newcommand{\eea}{\end{eqnarray}}

\begin{document}

\thispagestyle{empty}

\setcounter{page}{0}

\mbox{}
\vspace{-10mm}

\begin{center} {\bf \Large  Supersymmetric electric-magnetic duality as a manifest symmetry of the action for 
super-Maxwell theory and linearized supergravity}

\vspace{1.6cm}

Claudio Bunster$^{1,2}$ and Marc Henneaux$^{1,3}$

\footnotesize
\vspace{.6 cm}

${}^1${\em Centro de Estudios Cient\'{\i}ficos (CECs), Casilla 1469, Valdivia, Chile}

\vspace{.1cm}

${}^2${\em Universidad Andr\'es Bello, Av. Rep\'ublica 440, Santiago, Chile}

\vspace{.1cm}

${}^3${\em Universit\'e Libre de Bruxelles and International Solvay Institutes, ULB-Campus Plaine CP231, B-1050 Brussels, Belgium} \\

\vspace {15mm}

\end{center}
\centerline{\bf Abstract}
\vspace{.6cm}
{}For the free massless spin-one and spin-two field theories one may write the action in a form which is manifestly invariant under electric-magnetic duality.  This is achieved by introducing new potentials through solving the constraints of the Hamiltonian formulation.  The price for making electric-magnetic duality invariance manifest through this direct procedure is losing manifest Lorentz invariance. Both theories admit supersymmetric extensions, which make the bosonic fields and their corresponding fermionic partners to be parts of the same geometrical object, a supermultiplet.  We present in this paper the supersymmetric extension of the manifestly electric-magnetic duality invariant actions for the photon and the photino; and for the graviton and the gravitino. In each case the spinor fields transform under electric-magnetic duality in a chiral manner. For the spin-tree-half field, which possesses a gauge invariance, it is necessary to bring in a spinor ``prepotential".  As in previous cases the introduction of additional potentials to solve the constraints increases the number of gauge invariances of the action, thus keeping the number of degrees of freedom unaltered.  The similarity in the formulations for the photon-photino and graviton-gravitino systems is remarkable.

\vspace{.8cm}
\noindent

\newpage

\section{Introduction}
\setcounter{equation}{0}
Electric-magnetic duality is a fascinating symmetry that keeps appearing in many contexts. It was shown long ago \cite{Deser:1976iy} that for Maxwell theory it can be implemented as a manifest symmetry of the action in terms of {\em two} vector potentials, the ordinary one for the magnetic field and an additional one for the electric field.  The second potential was introduced by solving the constraint (Gauss' law) of the Hamiltonian formulation.  The price for achieving manifest electric-magnetic duality invariance through this direct procedure is losing manifest Lorentz invariance. 

The approach to electric-magnetic duality based on solving the Hamiltonian constraints has been fruitfully applied later on to $p$-forms, including Chern-Simons couplings and coupling to scalar fields defined on a coset space \cite{Bunster:2011aw,Bunster:2011qp,Comments}.  Furthermore it has been shown \cite{Henneaux:2004jw} that linearized Einstein gravity can also be formulated through this procedure in a manner which makes the action manifestly invariant under electric-magnetic duality transformations. That result was achieved by solving the Hamiltonian and momentum constraints in terms of two ``prepotentials"  which are two-index symmetric tensors. In all cases the introduction of additional potentials to solve the constraints increases the number of gauge invariances of the action, thus keeping the number of physical degrees of freedom unaltered.  

We consider in this paper the supersymmetric extensions of Maxwell theory and linearized Einstein gravity, i.e., super-Maxwell theory  and linearized supergravity.  We provide a formulation in which electric-magnetic duality is a manifest symmetry of the action for both the photon-photino and graviton-gravitino systems. For the gravitino one must introduce a new spinor ``prepotential" which is the fermionic superpartner of the symmetric tensor bosonic prepotentials previously found in linearized gravity. For the photino, since there are no fermionic gauge symmetries, and thus no fermionic constraints to solve, the original spinor field may be regarded as its own prepotential. In order to properly implement electric-magnetic duality together with supersymmetry it is necessary to define the former as acting chirally on the spinors. (The interplay between chirality and duality invariance was particularly stressed in \cite{Deser:1977ur}.) The very close parallel between the formulation for the photon-photino case and that for the graviton-gravitino case  is quite remarkable.  

The paper is organized as follows: Section \ref{SecII} reviews the supersymmetric extension of the standard one-potential Maxwell theory.  The supersymmetric extension of the manifestly electric-magnetic duality invariant two-potential formulation is then derived. It is shown that if one demands that electric-magnetic duality and supersymmetry commute,  then the duality transformation of the photino is a chiral rotation.   Section \ref{SecIII} employs the insights obtained in section \ref{SecII}  to develop the corresponding formulation for linearized supergravity.   Since the key results are remarkably similar to those of section \ref{SecII}, their proof is relegated to the Appendices in order not to deviate the attention of the reader from the main line of argument.  Finally section \ref{SecIV} is devoted to concluding remarks.   We work in four-dimensional Minkowski space throughout.

\section{Electric-Magnetic Duality for the Photon and the Photino}
\label{SecII}
\subsection{Two-potential formulation of the Maxwell theory}
If, besides the standard ``magnetic" vector potential defined through,
$$
\vec{B} \equiv \vec{B}_1 = \vec{\nabla} \times \vec{A}_1,
$$
one introduces an additional vector potential $\vec{A}_2$ through,
$$
\vec{E} \equiv \vec{B}_2 = \vec{\nabla} \times \vec{A}_2,
$$
one may rewrite the standard Maxwell action
\be
I = - \frac{1}{4} \int d^4x F^{\mu \nu} F_{\mu \nu} \label{Maxwell}
\ee
with,
$$
F_{\mu \nu} = \partial_\mu A_\nu - \partial_\nu A_\mu,
$$ 
in terms of the two potentials $A^a$ as \cite{Deser:1976iy} (see also \cite{DGHT})
\be
I = \frac{1}{2} \int dx^0 d^3x \left( \ep_{ab} \vec{B}^a \cdot \dot{\vec{A}}^b - \de_{ab} \vec{B}^a \cdot \vec{B}^b \right). \label{TwoPotential}
\ee
Here, $\ep_{ab}$ is given by $\ep_{ab} = - \ep_{ba}$, $\ep_{12} = +1$, or as a matrix
\be
\ep = \begin{pmatrix} 0 & 1 \\ -1 & 0 \end{pmatrix}, \; \; \; \; \ep^2 = - I  \label{Epsilon0}
\ee
where $I$ is the $2$ by $2$ unit matrix. 

The action (\ref{TwoPotential}) is invariant under rotations in the $(1,2)$ plane (``electric-magnetic duality rotations") ,
\be
\begin{pmatrix} \vec{A}^1 \\ \vec{A}^2  \end{pmatrix} \equiv \vec{\mathbf{A}} \; \;  \longrightarrow \; \; e^{\alpha \ep} \vec{\mathbf{A}} \label{E-MDuality}
\ee
because $\ep_{ab}$ and $\de_{ab}$ are invariant tensors.  Note that $\ep$ given by (\ref{Epsilon0}) generates {\em clockwise rotations.}  This convention makes the formulas below more symmetric.
The action (\ref{TwoPotential}) is also invariant under the gauge transformations,
$$
\vec{A}^a \; \;  \longrightarrow \; \; \vec{A}^a + \vec{\nabla} \Lambda^a.
$$

\subsection{Supersymmetric extension of the two-potential theory}
\label{SubsecIIB}
The action for the free supermultiplet formed by the spin one photon and the spin one-half photino has the form,
\be
I = I_{\textrm{BOSE}} + I_{\textrm{FERMI}} \label{BoseFermi},
\ee
where $I_{\textrm{BOSE}}$ is the action for the photon and $I_{\textrm{FERMI}}$ is given by,
$$
I_{\textrm{FERMI}} = - \frac{i}{2} \int d^4 x \, \bar{\psi} \ga^\mu \partial_\mu \psi
$$
where $\psi$ is an anticommuting Majorana spinor. The action $I_{\textrm{FERMI}}$ is invariant under the chirality transformation,
$$
\psi \; \; \longrightarrow \; \; e^{\beta \ga_5} \psi,
$$
which is an $SO(2)$-rotation because,
$$
(\ga_5)^2 = - I.
$$
The matrix $I$ is here the $4$ by $4$ unit matrix.

If one takes for $I_{\textrm{BOSE}}$ in (\ref{BoseFermi}) the standard one-potential action (\ref{Maxwell}), the sum $I_{\textrm{BOSE}} + I_{\textrm{FERMI}}$ is invariant under the infinitesimal supersymmetry transformations
\bea
&& \de A_\mu = i \bar{\ep} \ga_\mu \psi, \label{Super1,a}\\
&& \de \psi = \frac{1}{2} \ga^{\mu \nu} F_{\mu \nu} \ep,  \label{Super1,b}
\eea
where $\ep$ is a constant anticommuting Majorana spinor. [See \cite{VanHolten} for a lucid presentation.] In order to determine the realization of the supersymmetry transformation in the two-potential theory, we observe that starting from it one can go to a purely electric or a purely magnetic representation by eliminating either the electric or the magnetic potential respectively. (We follow standard terminology: the electric representation is the one where the electric charge couples to the usual ``magnetic" vector potential.) In the magnetic representation, the Maxwell action takes the form (\ref{Maxwell}) with $F_{\mu \nu}$ replaced by the field strength of the second vector potential $A^2_\mu$,
$$
F^2_{\mu \nu}  = \partial_\mu A^2_\nu - \partial_\nu A^2_\mu.
$$
Therefore, the action is invariant under the transformation 
\bea
&& \de A^2_\mu = i \bar{\eta} \ga_\mu \psi, \label{Super2,a}\\
&& \de \psi = \frac{1}{2} \ga^{\mu \nu} \,   F^2_{\mu \nu} \eta, \label{Super2,b}
\eea
for some spinor $\eta$.
To determine the relation between $\eta$ and $\ep$, one recalls that on-shell one has,
$
F^2_{\mu \nu} =  \,  ^*\hspace{-.06cm}F_{\mu \nu} \equiv \frac{1}{2} \ep_{\mu \nu \rho \sigma} F^{\rho \sigma} %\label{Hodge}
$, and one uses the identity $   ^*\hspace{-.06cm} \ga_{\mu \nu} = \ga_5 \ga_{\mu \nu}$.  This gives  $\eta = - \ga_5 \ep$.  Thus the supersymmetry transformation in the magnetic representation is given by
\bea
&& \de A^2_\mu = i \bar{\ep} \ga_\mu \ga_5 \psi. \label{Super2,c}\\
&& \de \psi = -\frac{1}{2} \ga^{\mu \nu} \,   F^2_{\mu \nu} \ga_5 \ep, \label{Super2,d}
\eea
The presence of $\ga_5$ makes $\de A^2_\mu$ a pseudo-vector, as it should be the case since $A^2_\mu$ is the potential for $^*F_{\mu \nu}$.

To pass to the two-potential representation, we may start from the electric representation and  expand the sum over $[\mu \nu]$ on the right-hand side of (\ref{Super1,b}) into electric  [0i] and magnetic [mn] pieces.  Inserting the expression for the electric and magnetic fields as curls of their respective potentials, we obtain,
\be
\de \psi = \ga^{rs} \partial_r \left( A^1_s - \ga_5 A^2_s \right). \label{Super3,b}
\ee
This expression, together with the spatial parts of (\ref{Super1,a}) and (\ref{Super2,c}),
\be
\de \vec{{\mathbf A}} = \de \begin{pmatrix} \vec{A}^1 \\ \vec{A}^2 \end{pmatrix} = i \bar{\ep} \vec{\ga} \begin{pmatrix} 1 \\ \ga_5 \end{pmatrix} \psi, \label{Super3,a}
\ee
is the the supersymmetry transformation, which leaves invariant the action (\ref{BoseFermi}) with $I_{\textrm{BOSE}}$ given by (\ref{TwoPotential}).

Now we pass to discuss the interplay between supersymmetry and duality.  To that effect we observe that under an electric-magnetic duality transformation of the form (\ref{E-MDuality}) on $\vec{A}^a$, one has,
$$
\vec{A}^1 - \ga_5 \vec{A}^2 \; \; \longrightarrow \; \; e^{\al \ga_5} \left( \vec{A}^1 - \ga_5 \vec{A}^2 \right).
$$
Then $\de \psi$ in (\ref{Super3,b}) transforms as,
$$
\de \psi \; \;  \longrightarrow \; \;  e^{\al \ga_5} \de \psi.
$$
Conversely, if one transforms $\psi$ as,
$$
\psi \; \;  \longrightarrow \; \;  e^{\al \ga_5}  \psi,
$$
then $\delta \vec{{\mathbf A}}$ in (\ref{Super3,a}) transforms according to a duality rotation,
$$
\de \vec{{\mathbf A}} \; \; \longrightarrow \; \; e^{\al \ep} \de \vec{{\mathbf A}}.
$$
This implies that {\em supersymmetry and electric-magnetic duality commute} if one defines the latter as a transformation of {\em both the vectors $\vec{A}^a$ and the spinor $\psi$},
\be
\vec{\mathbf{A}} \; \;  \longrightarrow \; \; e^{\alpha \ep} \vec{\mathbf{A}}, \; \; \; \; \;  \psi \; \;  \longrightarrow \; \;  e^{\al \ga_5}  \psi. \label{SuperDuality}
\ee
The parallelism between these two transformations becomes all the more poignant if one uses the Majorana representation given in the Appendix \ref{Conventions}.  Then the Majorana spinors are real and
$$
\ga_5 = \begin{pmatrix} 0 & I \\ - I & 0 \end{pmatrix}. 
$$ On the other hand, as given above,
$$
\ep = \begin{pmatrix} 0 & 1 \\ -1 & 0 \end{pmatrix}.
$$
Actually $\ep$ can be thought of as a $4 \times 4$ matrix when acting on the four independent components of the gauge invariant curl $\vec{\nabla} \times \vec{A}$.  It then becomes $\ga_5$ itself!

The beautiful similarity between the duality transformation properties of the vector and spinor fields arose because although uncoupled they were related by supersymmetry. 

The supersymmetry transformation of the non-manifestly Lorentz invariant two potential formulation of the super-Maxwell theory can also be obtained from the results of \cite{Hillmann:2009zf} on the manifestly $E_{7,7}$-invariant formulation of maximal supergravity in four space-time dimensions by truncating the corresponding formulas to a subsector of the theory containing only one vector field and one of its supersymmetric spin 1/2 partner.  We have provided above a different, direct derivation of the supersymmetry transformations using the existence of the  ``electric" and ``magnetic" representations.

\section{Electric-Magnetic Duality for the Graviton and the Gravitino}
\setcounter{equation}{0}
\label{SecIII}
We will develop in this section the formulation of linearized supergravity in terms of prepotentials,  which makes electric-magnetic duality to be a manifest invariance of the action.  The corresponding formulation for linearized gravity alone was given in \cite{Henneaux:2004jw}. In that case the solution of the Hamiltonian and momentum constraints, which generate the Hamiltonian version of linearized spacetime diffeomorphisms, led to the introduction of two symmetric tensor prepotentials for the canonical pair formed by the spatial metric and its canonical conjugate, the extrinsic curvature.

Supergravity brings in an additional fermionic gauge invariance and, with it, an additional Majorana spinor constraint that generates it.  That constraint can be solved to express the canonically self-conjugate vector-spinor gauge potential of the theory in terms of a vector-spinor prepotential.

It was natural to expect, by analogy with the case of the photon and the photino, that a similar formulation for the graviton and the graviton, which exhibits manifest electric-magnetic duality invariance, could be achieved by focusing on the bosonic and fermionic prepotentials and their transformation properties under global supersymmetry.  This not only turned out to be the case, but the result transcended the most optimistic expectations: the ensuing formulation for the spin two/spin three-half multiplet is not just similar but {\em identical} to the one for the spin one/spin one-haf case.  For that reason, we will only give the results in the main text and provide the proofs of the key equations in the Appendices.

\subsection{Prepotentials for the graviton and the gravitino}
For linearized gravity the fields appearing in the canonical formulation are the linearized spatial metric $h_{ij}$ ($g_{ij} = \de_{ij} + h_{ij}$) and its canonically conjugate momentum $\pi^{ij}$.  They obey the Hamiltonian and momentum constraints,
\bea
&& \partial^m \partial^n h_{mn} - \triangle h = 0, \nonumber \\
&& \pi^{mn}_{\; \; \; \; \; , n} = 0, \nonumber
\eea
where $h = h^m_{\; \; m}$ and $\triangle$ is the Laplacian. 

One may solve \cite{Henneaux:2004jw} these constraints  by introducing two prepotentials $Z^a_{mn}$ such that,
\bea
&& \pi^{mn} = \ep^{mpq} \ep^{nrs} \partial_p \partial_r Z^1_{qs}, \nonumber \\
&& h_{mn} = \ep_m^{\; \; rs} \partial_r Z^2_{sn} + \ep_n^{\; \; rs} \partial_r Z^2_{sm} . \nonumber
\eea
Note that $Z^1$ transforms as a scalar under inversions, whereas $Z^2$ transforms as a pseudo-scalar, just as in the Maxwell case. Given $h_{mn}$ and $\pi^{mn}$ up to a gauge transformation, there are ambiguities
\be
Z^a_{mn} \; \; \longrightarrow \; \; Z^a_{mn} + \partial_m \xi^a_n + \partial_n \xi^a_m + \xi^a  \de_{mn}, \label{GaugeBose}
\ee
which constitute the gauge transformations of the prepotentials.  The terms with $\xi^a_m$ are linearized diffeomorphisms, while the last term takes the form of a linearized Weyl rescaling.

In linearized supergravity there is, in addition to the conjugate pair $(h_{mn},\pi^{mn})$, the self-conjugate Majorana vector-spinor $\psi_m$, which is an anticommuting field that has four real components in a Majorana representation, in which the Dirac matrices are real.

The vector-spinor $\psi_m$ obeys the constraint
\be
\ga^{mn} \partial_m \psi_n = 0 \label{SpinorConstraint0}
\ee
which may be solved (see Appendix \ref{SolutionPsi}) in terms of another vector-spinor $\chi_p$ as,
\be
\psi_m = \ga_5 (\de_{ms} - \ga_{ms} ) \ep^{skp} \partial_k \chi_p. \label{SpinorPrepotential}
\ee
Given $\psi_m$ up to a gauge transformation $\psi_m \rightarrow \psi_m + \partial_m \ep$, there are the ambiguities,
\be
\chi_p \; \; \longrightarrow \; \; \chi_p + \partial_p \eta + \ga_p \rho, \label{GaugeFermi}
\ee
where $\eta$, $\rho$ are Majorana spinors.  These constitute the gauge symmetries for the fermionic prepotential $\chi_p$.  Note the close resemblance of the transformations for the twelve $Z^a_{mn}$ and the twelve $\chi^A_m$.

It is noteworthy that, while the gauge transformations of the bosonic prepotentials are those of linearized conformal gravity, the gauge transformations (\ref{GaugeFermi}) of the fermionic prepotential turn out to be exactly those of linearized conformal supergravity, with two independent fermionic symmetries \cite{Kaku:1977pa}, parametrized here by the two independent spinor parameters $\eta$ and $\rho$.

\subsection{Supersymmetry transformations of the prepotentials}

First, we recall how to infer the global supersymmetry transformations of the free graviton and gravitino fields $h_{\mu \nu}$ and $\psi^A_\mu$.  Starting from them we will obtain the corresponding transformations  for the prepotentials $Z^a_{mn}$ and $\chi^A_m$.  Actually, the procedure is nothing but undoing the steps that originally led to supergravity through gauging the global symmetries of the free spin-two/spin-three-half theory.  We follow this ``reversed route" because it brings in immediately the geometrical quantities that we need, i.e., the spatial metric and connection, and the extrinsic curvature.  

One starts by imagining expanding the action for full supergravity  in powers of the fields and their derivatives as $I^{(2)} +I^{(3)} + \cdots$  where $I^{(k)}$ is of $k$-th polynomial degree. The quadratic part $I^{(2)}$ is the action for linearized supergravity and is the sum of the free spin-2 and free spin-3/2 actions.
Similarly, one can expand the supersymmetry transformations under which the full action is invariant as $\de_\ep \Phi^i = \de^{(1)}_\ep \Phi^i + \de^{(2)}_\ep \Phi^i + \cdots$ where the expansion is again performed according to the polynomial degree, counting also the gauge parameter $\ep$.  Here, $\Phi^i$ stands for all the fields. Explicitly,
\be  \de^{(1)}_\ep  h_{\mu \nu} = 0, \; \;  \de_\ep^{(2)} h_{\mu \nu}=i \left( \bar{\ep} \ga_\mu \psi_\nu +  \bar{\ep} \ga_\nu \psi_\mu\right) \label{hBoson} \ee
for the graviton and
\be \de^{(1)}_\ep  \psi_\mu = 4\partial_\mu \ep, \; \;  \de_\ep^{(2)} \psi_\mu = -  \omega_{\mu \rho \sigma} \ga^{\rho \sigma} \ep, \label{psiFermion} \ee
for the gravitino.  Here, $\omega_{\mu \rho \sigma}$ is the linearized spin connection (see appendix \ref{Conventions}) so that $ \de^{(1)}_\ep  \psi_\mu + \de_\ep^{(2)} \psi_\mu = 4 D_\mu \ep$.  (We have chosen a normalization so that the formulas for the prepotentials look simple.)

The invariance of the full action under the full supersymmetry transformations implies, when expanded according to the polynomial degree, a chain of equations: $\de^{(1)}_\ep I^{(2)} = 0$ (lowest order), $\de^{(2)}_\ep I^{(2)} + \de^{(1)}_\ep I^{(3)} = 0$ (next order), etc. The first of these equations just expresses the invariance of the action $I^{(2)}$ of linearized supergravity (free Rarita-Schwinger action) under the abelian gauge supersymmetry transformations $ \de^{(1)}_\ep \Phi^i$.   As we have just recalled above, these transformatione reduce  to $\de^{(1)}_\ep \psi_\mu = 4 \partial_\mu \ep$, $\de^{(1)}_\ep h_{\mu \nu} = 0$. 

Since $ \de^{(1)}_\ep \Phi^i$ contains only derivatives of $\ep$,   it is identically equal to zero for constant (i.e., spacetime independent) $\ep$'s.  Therefore,   $\de^{(1)}_\ep I^{(k)} \equiv 0$ for all $k$'s, and in particular  $\de^{(1)}_\ep I^{(3)} \equiv 0$, when $\ep$ is taken to be a constant. It follows from $\de^{(2)}_\ep I^{(2)} + \de^{(1)}_\ep I^{(3)} = 0$  that the action $I^{(2)}$ of linearized supergravity  is invariant under the rigid supersymmetry transformations $ \de^{(2)}_\ep \Phi^i$ with constant $\ep$'s, $\de^{(2)}_\ep I^{(2)}= 0$.  So, in addition to the gauge invariance under the local supersymmetry transformations $\de^{(1)}_\ep \Phi^i$ (with $\ep$ an arbitrary spacetime dependent spinor), the free action of linearized supergravity possesses also the invariance under the ``rigid supersymmetry"  $ \de^{(2)}_\ep \Phi^i$ with constant $\ep$ .  It is this latter transformation that we want to write for the prepotentials.

The derivation is given in Appendix \ref{SusyPrepotentials}. One finds, starting from (\ref{hBoson}) and (\ref{psiFermion}) with constant $\ep$, 
\be
\de \chi_p = \ga^{rs} \partial_s \left( Z^1_{\, rp} - \ga_5 Z^2_{\, rp} \right) \ep, \label{Super4,b}
\ee
and,
\be
\de {\mathbf Z}_{mn} = \de \begin{pmatrix} Z^1_{\, mn} \\ Z^2_{mn} \end{pmatrix} = i \bar{\ep} \left( \ga_m \begin{pmatrix} 1 \\ \ga_5 \end{pmatrix} \chi_n +   \ga_n \begin{pmatrix} 1 \\ \ga_5 \end{pmatrix} \chi_m    \right) .  \label{Super4,a}
\ee
Equations (\ref{Super4,b}) and  (\ref{Super4,a}) are the same as (\ref{Super3,b}) and  (\ref{Super3,a}) with ${\mathbf A}_m$ replaced by ${\mathbf Z}_{mn}$ and $\chi$ replaced by $\chi_p$.  Therefore, all the conclusions of Subsection \ref{SubsecIIB} translate literally. Electric-magnetic duality acts on both ${\mathbf Z}$ and $\chi$ as,
$$
{\mathbf Z }\; \; \;  \longrightarrow \; \; \; e^{\al \ep}\,{\mathbf Z } ; \; \; \; \; \; \;  \chi \; \; \; \longrightarrow \; \; \; e^{\al \ga_5} \chi,
$$
and it commutes with supersymmetry.

Note in this context that, unlike the photon-photino case, if the $Z^a_{\, sp}$ in (\ref{Super4,b}) undergo a gauge transformation, the right hand side of that equation experiences a transformation of the form (\ref{GaugeFermi}). Similarly for (\ref{Super4,a}).  This phenomenon does not have an analog for the spin one - spin one-half case because there the spin one-half field is its own prepotential and does not possess gauge freedom.

\subsection{Manifestly duality invariant action}
To complete the presentation for the graviton case, we present here the action expressed in terms of prepotentials. 

The action is,
$$
I = I_{\textrm{BOSE}} + I_{\textrm{FERMI}},
$$
where $I_{\textrm{BOSE}}$ was given in \cite{Henneaux:2004jw},
\be
S[Z_a^{\; mn}] = \int dt \left[ \int d^3x \, \ep^{ab} \ep^{mrs} \left(\partial^p \partial^q \partial_r Z_{aps} - \triangle \partial_r Z_{a\;   s}^{\;q}\right) \dot{Z}_{bqm} - H_{\textrm{BOSE}} \right], \label{Action0}
\ee
with,
\bea
H_{\textrm{BOSE}} &=& \int d^3x\,  \de^{ab} \left( \triangle Z_{aij} \, \triangle Z_b^{\; ij} + \frac{1}{2} \partial^k \partial^m Z_{akm} \partial^q\partial^n Z_{bqn} + \partial^k \partial^m Z_{akm} \triangle Z_{b}\right) \nonumber \\
&& +  \int d^3x \, \de^{ab} \left(-2 \partial_m \partial_i Z_{a}^{\; ij} \partial^m\partial^k Z_{bkj} - \frac{1}{2} \triangle Z_{a} \triangle Z_{b}\right),
\eea
and $Z^a = Z^{am}_{\; \; \; \; m}$.
This action is manifestly invariant under electric-magnetic duality because $\ep_{ab}$ and $\de_{ab}$ are invariant tensors. A more transparent rewritteng of the action $I_{\textrm{BOSE}}$ in which its gauge invariance under linearized differomorphisms and Weyl rescalings are made more explicit is given in \cite{BHHToAppear}.

The fermionic action is obtained by inserting in the Rarita-Schwinger action the expression (\ref{SpinorPrepotential}) for the original vector-spinor $\psi_m$ in terms of its prepotential $\chi_p$.  The component $\psi_0$ drops out because the spinor constraint vanishes identically in terms of the prepotential. 

One finds (see Appendix \ref{RRAction2}),
$$
I_{\textrm{FERMI}} =  \int dx^0 \left(- i \int d^3 x \bar{\Sigma}_{mn} \ga^{mnp} \dot{\chi}_p  - H_{\textrm{FERMI}} \right),
$$
with
$$
H_{\textrm{FERMI}} = -\frac{i}{4} \int d^3x \, \bar{\Sigma}^{mn} \ga^0 \ga_5 ( \delta_{mk} - 2 \ga_{mk}) \Delta_n^{\; \; k} .
$$
Here the tensor-spinor field strengths $\Delta_{pq}$ and $\Sigma_{pq}$ are respectively defined by 
$$\Delta_{pq} =\partial_p \chi_q - \partial_q \chi_p, $$
and
$$\Sigma_{pq} = \frac{1}{2} \ga_5 (\de_{qs} - \ga_{qs} )\ep^{skm} \partial_p \Delta_{km} - (p \leftrightarrow q).
$$
The tensor-spinor $\Sigma_{pq}$ is gauge invariant under all gauge symmetries of $\chi_p$ and fulfills the identity $\Sigma_{pq} \ga^{pq} = 0$, in addition to $\partial_{[r}\Sigma_{pq]}= 0$.

The action $I_{\textrm{FERMI}}$ is invariant under the chirality transformation $\chi_p \rightarrow e^{\al \ga_5} \chi_p$ and is therefore electric-magnetic duality invariant.

\section{Concluding Remarks}
\label{SecIV}

In this paper, we have derived the manifestly electric-magnetic duality invariant formulation of super-Maxwell theory and linearized supergravity. The manifestly duality invariant action of super-Maxwell theory involves two potentials, while the manifestly duality invariant action of linearized supergravity involves one fermionic prepotential besides the two bosonic prepotentials of linearized gravity \cite{Henneaux:2004jw}.  The supersymmetry transformations of the two potentials and the spin-one-half field (super-Maxwell theory) and of the three prepotentials (linearized supergravity) have been explicitly written and are local.  They take a remarkably similar form in both cases.

Supergravity and electric-magnetic duality have a well known fruitful interplay \cite{Ferrara:1976iq}.  But we stress that the duality discussed here in the context of linearized supergravity is a gravitational duality that acts on the graviton.   It is present already for the $N=1$ theory (and for that matter, even for $N=0$), without the vector or scalar fields present in the extended models and on which duality is traditionally considered. 

It would be of interest to include a cosmological constant in our analysis and to and analyze  gauged supergravities.  Existing work in that direction is encouraging \cite{Julia:2005ze} (see also \cite{Deser:2004xt} for related comments). 
  
{}Finally, the asymptotic properties of the manifestly duality invariant formulation of supergravity are also of definite interest, given the enlargement of the gauge group.  This problem is currently under study (see \cite{Argurio:2008zt} in that context).

\section*{Acknowledgments} 
Both authors  thank  the Alexander von Humboldt Foundation for Humboldt Research Awards.  The work of M.H. is partially supported by the ERC through the ``SyDuGraM" Advanced Grant, by IISN - Belgium (conventions 4.4511.06 and 4.4514.08) and by the ``Communaut\'e Fran\c{c}aise de Belgique" through the ARC program.  The Centro de Estudios Cient\'{\i}ficos (CECS) is funded by the Chilean Government through the Centers of Excellence Base Financing Program of Conicyt.    

\break

\noindent
{\bf \Large{Appendices}}

\appendix

\section{Conventions - $\gamma$-matrices}
\label{Conventions}
\setcounter{equation}{0}

The Dirac $\gamma$-matrices fulfill,
$$
\gamma_\mu \gamma_\nu + \gamma_\nu \gamma_\mu  = 2 \eta_{\mu \nu}
$$
where $\eta_{\mu \nu}$ has the ``mostly $+$" signature $(-, +, +, +)$.

We adopt a Majorana representation where the $\gamma$-matrices are real, with antisymmetric $\gamma_0$ and symmetric $\gamma_k$'s,
$$
(\gamma_0)^T = - \gamma_0, \; \; \; \; \;  (\gamma_k)^T = \gamma_k.
$$

The matrix $\gamma_5$ is defined through,
$$
\gamma_5 = \gamma_0 \gamma_1 \gamma_2 \gamma_3 ,
$$
and fulfills,
$$
(\gamma_5)^2 = - I, \; \; \; \; (\gamma_5)^T = - \gamma_5.
$$

We define $\gamma_{\mu \nu} = \frac{1}{2}(\gamma_\mu \gamma_\nu - \gamma_\nu \gamma_\mu)$ and $\gamma_{\mu \nu \rho} = \frac{1}{3!}(\gamma_\mu \gamma_\nu \gamma_\rho + \gamma_\nu \gamma_\rho \gamma_\mu + \gamma_\rho \gamma_\mu \gamma_\nu - \gamma_\mu \gamma_\rho \gamma_\nu - \gamma_\rho \gamma_\nu \gamma_\mu - \gamma_\nu \gamma_\mu \gamma_\rho) = \epsilon_{\mu \nu \rho \sigma} \gamma_5 \gamma^\sigma$ with $\epsilon_{0123} = +1$.

Useful relations are:
$$ \epsilon^{kmp}\gamma_p = \gamma^{km} \gamma_0 \gamma_5, $$
$$\gamma^{mn} \gamma^{s} = - \epsilon^{mns} \gamma_0 \gamma_5 + \delta^{ns} \gamma^{m} -  \delta^{ms} \gamma^{n}, $$
$$\ga^{mn} \ga^{rs} = \delta^{nr} \ga^{ms} -\delta^{ns} \ga^{mr} - \delta^{mr} \ga^{ns} + \delta^{ms} \ga^{nr} + \delta^{nr} \de^{ms} -  \delta^{mr} \de^{ns},$$
$$ \gamma^{mn} \gamma_{np} = 2 \delta^m_{\; p} + \gamma^m_{\; \; \; p} ,$$
$$  \gamma^{mn} \gamma^{0s} = - \epsilon^{mns} \gamma_5 + \delta^{ns} \gamma^{0m} -  \delta^{ms} \gamma^{0n},$$
$$ \ga^{abc} = \ep^{abc} \ga^0 \ga_5, $$
$$^* \ga^{\mu \nu} = \ga_5 \ga^{\mu \nu}. $$

An explicit Majorana represention of the $\ga$-matrices is given by, 
$$
\ga_0 = \begin{pmatrix} 0 & 1 & 0 & 0 \\ -1 & 0 & 0 & 0 \\ 0 & 0 & 0 & -1 \\ 0 & 0 & 1 & 0 \end{pmatrix}, \; \; \; 
\ga_1 = \begin{pmatrix} 1 & 0 & 0 & 0 \\ 0 & -1 & 0 & 0 \\ 0 & 0 &-1 & 0 \\ 0 & 0 & 0 & 1 \end{pmatrix},
$$
$$
\ga_2 = \begin{pmatrix} 0 & 0 & -1 & 0 \\ 0 & 0 & 0 & -1 \\ -1 & 0 & 0 & 0 \\ 0 & -1 & 0 & 0 \end{pmatrix}, \; \; \; 
\ga_3 = \begin{pmatrix} 0 & -1 & 0 & 0 \\ -1 & 0 & 0 & 0 \\ 0 & 0 &0 & 1 \\ 0 & 0 & 1 & 0 \end{pmatrix},
$$
$$
\ga_5 = \begin{pmatrix} 0 & 0 & 1 & 0 \\ 0 & 0 & 0 & 1 \\ -1 & 0 & 0 & 0 \\ 0 & -1 & 0 & 0 \end{pmatrix}.
$$
The Dirac adjoint is defined by
$$
\bar{\psi} = \psi^T \ga_0
$$

The covariant derivatives of a spinor field is,
$$
D_\mu \psi = \partial_\mu \psi - \frac{1}{4} \omega_{\mu \rho \sigma} \ga^{\rho \sigma} \psi.
$$
The spin connection in the linearized theory is,
\be
\omega_{\mu \rho \sigma} = \frac{1}{2} \left( \partial_\rho h_{\mu \sigma} - \partial_\sigma  h_{\mu \rho} \right). \label{LinearizedConnection}
\ee

\section{Action for the Rarita-Schwinger field}
\label{RRAction}
\setcounter{equation}{0}

The one-potential action for the Rarita Schwinger field is,
$$
I_{\textrm{FERMI}} = - \frac{i}{2} \int d^4x \bar{\psi}_\al \ga^{\al \beta \ga} \partial_\beta \psi_\ga.
$$
The equations of motion for the spin $\frac{3}{2}$ field are, 
$$
\gamma^{\al \beta \ga} \left( \partial_\beta \psi_\ga - \partial_\ga \psi_\beta \right) = 0.
$$
{}For $\al = 0$, one obtains the constraint
\be
\ga^{mn} \partial_m \psi_n = 0, \label{SpinorConstraint}
\ee
whereas for $\al = a$, 
\be
\ga^{ac}\left(\partial_0 \psi_c - \partial_c \psi_0 \right) = \ep^{abc} \ga_5 \partial_b \psi_c.
\ee
One can solve for $\partial_0 \psi_a$,
\be
 \partial_0 \psi_a = \partial_a \psi_0 - \frac{1}{2} \ep_{abc} \ga_5 \partial^b \psi^c + \frac{1}{2} \ga_{am} \ep^{mbc} \ga_5 \partial_b \psi_c .  \label{EoMRS}
\ee

We also recall the expression of the linearized extrinsic curvature,
\be
K_{ij} = - \frac{1}{2} \left(\partial_0 h_{ij} - \partial_i h_{0j} - \partial_j h_{0i} \right), \label{Extrinsic}
\ee
and that the ``first Hamiltonian equation of motion" for the spin-2 field may be written ,
$$
\pi^{ij} = - K^{ij} + K \delta^{ij}.
$$

\subsection{Solving the constraint for $\psi_k$}
\label{SolutionPsi}
The constraint (\ref{SpinorConstraint}) may be rewritten as,
$$
\gamma^{mn}  \psi_n = 2 \partial_k \left(\epsilon^{kmp} \ga_5 \chi_p \right),
$$
for some ``prepotential"  $\chi_p$, which is also a vector-spinor.  
One can explicitly express $\psi_k$ in terms of $\chi_k$,
\begin{equation}
\psi_m =   \ga_5 \left(\de_{ms} - \ga_{ms} \right)\epsilon^{skp}  \partial_k \chi_p . \label{SolForPsi}
\end{equation}

Given $\psi_p$ up to a gauge transformation $\psi_p \rightarrow \psi_p + \partial_p \varepsilon$, $\chi_p$ is determined up to
\begin{equation}
\chi_p \rightarrow \chi_p + \partial_p \eta + \gamma_p \rho,
\label{GaugeForChi} \end{equation}
where $\eta$ and $\rho$  are spinor fields.   Note that $\varepsilon = 2 \gamma_0 \rho$ and that $\psi_m$ is invariant if $\chi_p \rightarrow \chi_p+ \partial_p \eta$.

As stressed in the main text,  it is remarkable that (\ref{GaugeForChi}) are just the linearized fermionic gauge transformations of conformal gravity which, as it is well known, has two independent supersymmetries (``$Q$" and ``$S$") \cite{Kaku:1977pa}.

\subsection{Action of linearized supergravity in terms of the prepotentials}
\label{RRAction2}
To write the Rarita-Schwinger action in terms of the vector-spinor $\chi_p$, we first introduce some tensor-spinors.  These are the field strength of $\chi_p$,
$$\Delta_{pq} = \partial_p \chi_q - \partial_q \chi_p$$
and the field strength of $\psi_p$,
$$\Sigma_{pq} = \partial_p \psi_q - \partial_q \psi_p$$
where $\psi_m$ is the function of $\chi_p$ defined by (\ref{SolForPsi}), which one may rewrite as,
$$\psi_m =   \frac{1}{2}\ga_5 \left(\de_{ms} - \ga_{ms} \right)\epsilon^{skp} \,  \Delta_{kp}.  $$
The field strength $\Delta_{pq}$ is invariant under the $\eta$-gauge symmetry in (\ref{GaugeForChi}) but not the $\rho$-one, while the tensor-spinor $\Sigma_{pq}$ is invariant under both.   Furthermore,
$\Sigma_{pq}$ fulfills the identity,
$$\ga^{pq} \Sigma_{pq} = 0.$$

If one replaces in the Rarita-Schwinger action the vector-spinor $\psi_p$ by its expression (\ref{SolForPsi}) in terms of $\chi_p$, one gets,
$$I_{\textrm{FERMI}} =  \int dx^0 \left(- i \int d^3 x \bar{\Sigma}_{mn} \ga^{mnp} \dot{\chi}_p  - H_{\textrm{FERMI}} \right),$$
with
$$ H_{\textrm{FERMI}} = -\frac{i}{4} \int d^3x \, \bar{\Sigma}^{mn} \ga^0 \ga_5 ( \delta_{mk} - 2 \ga_{mk}) \Delta_n^{\; \; k} .$$

\section{Supersymmetry transformations of the prepotentials}
\label{SusyPrepotentials}
\setcounter{equation}{0}

\subsection{Fermionic prepotential $\chi_p$}
One has,
$$
\delta \psi_m = 4 D_m \epsilon = -  \omega_{m \rho \sigma} \gamma^{\rho \sigma} \epsilon
$$
for constant supersymmetry transformations, where we have kept only the terms linear in the fields.  
There are two terms in this equation, one involving only the spatial spin connection $\omega_{m r s}$ and the other involving the extrinsic curvature, which is proportional to $ \omega_{m 0 s}$, as shown by  (\ref{LinearizedConnection}) and (\ref{Extrinsic}).  Explicitly,
$$
\delta \psi_m = \delta_h \psi_m + \delta_\pi \psi_m,
$$
where,
$$
\delta_h \psi_m = - \partial_r h_{ms} \ga^{rs} \epsilon,
$$
and
$$
 \delta_\pi \psi_m = -  (\partial_0 h_{ms} - \partial_s h_{m0})\ga^{0s} \epsilon.
$$
 
 \subsubsection{Term containing the prepotential $Z^1_{mn}$}
 Turn first to $\delta_\pi \psi_m$. By adding the gauge transformation   $\partial_m \left(  h_{s0} \ga^{0s} \epsilon \right)$  to $\psi_m$, which is permissible, one can rewrite it as,
$$
 \delta_\pi \psi_m =  2 K_{ms}\ga^{0s} \epsilon.
$$
 To evaluate the corresponding variation of the spinor prepotential $\chi_p$, we multiply this expression by $\ga^{mn}$ to get,
$$
\ga^{mn}  \delta_\pi \psi_n =  2 K_{ns} \ga^{mn}\ga^{0s} \epsilon.
 $$
 Using  the  identity for $\ga^{mn}\ga^{0s}$ given above, this can be transformed into, 
$$
\ga^{mn}  \delta_\pi \psi_n =  2 \pi^m_{\; \; \; n} \ga^{0n} \epsilon =  - 2 \pi^{mn} \ga_{0n} \epsilon.
$$

 On the other hand, $\ga^{mn}  \delta_\pi \psi_n = 2 \partial_k \left(\ep^{kmp} \ga_5 \de \chi_p \right)$, while $\pi^{mn} = \ep^{mkp} \ep^{nrs} \partial_k \partial_r P_{ps}$, so that we have,
 $$ \ep^{kmp}\partial_k \left(\ga_5  \de \chi_p-  \ep^{nrs} \partial_r P_{ps} \ga_{0n} \epsilon \right) = 0,$$ from which one infers, up to a gauge transformation $\chi_p \rightarrow \chi_p + \partial_p \eta$ that can be dropped,
 $$
\ga_5  \de \chi_p =  \ep^{nrs} \partial_r P_{ps} \ga_{0n} \epsilon,
 $$
an expression that can be rewritten as,
 \begin{equation}
 \de \chi_p =   \partial_s P_{pr} \ga_{5} \ga^{rs} \epsilon. \label{SUSY1ForChi}
 \end{equation}

 \subsubsection{Term containing the prepotential $Z^2_{mn}$}
 For the other term, $\delta_h \psi_m$,  one has,
$$
\ga^{mn} \delta_h \psi_m = - \partial_r h_{ms} \ga^{mn}\ga^{rs} \epsilon.
 $$
Using the  identity for the product $\ga^{mn}\ga^{rs}$ given above yields,
\bea
\ga^{mn} \delta_h \psi_m &=&  \left(-\partial^n h_{ns} \ga^{ms} + \ga^{mr} \partial_r h -\partial^r h_{n}^{\; \;  m} \ga^{nr} \right) \ep \nonumber \\
&& + \left(- \partial^n h_{mn} + \partial^m h \right) \ep. \nonumber
\eea
 Expressing $h_{ij}$ in  terms of $\Phi_{ij}$ gives,
\be
\ga^{mn} \delta_h \psi_m =  \ep^{kmp} \partial_k \left( \partial^r \Phi_{br} \ga_p^{\; \; b} + 2 \partial_b \Phi_{pc} \ga^{cb} - \partial_b \Phi \ga_p^{\; \; b} + \partial^n \Phi_{np} \right) \ep   \label{expression}
\ee

This expression takes the form $\ep^{kmp} \partial_k \Xi_p$ for some $\Xi_p$ and is at the same time equal to,
\be
2 \ep^{kmp} \ga_5 \partial_k \de_h \chi_p, \label{expression27}
\ee
if one expresses the variation of $\psi_m$ in terms of the variation of $\chi_p$.  Accordingly, by mere comparison of (\ref{expression27}) with (\ref{expression}), one can read off the supersymmetry variation of the vector-spinor prepotential.   Now, one can always add to  $\de \chi_p$ an arbitrarily chosen gauge transformation of $\chi_p$.  This is just a matter of definition and we will use this freedom to simplify the supersymmetry variation of $\chi_p$.

It turns out that the first term and the last term in the right-hand side of (\ref{expression}) are just a gauge transformation of $\chi_p$ since
\bea \partial^r \Phi_{br} \ga_p^{\; \; b} + \partial^n \Phi_{np} &=& - \frac{1}{2}\partial^r \Phi_{br} \ga^b \ga_p + \partial^n \Phi_{np} + \ga_p A \nonumber \\&=& - \partial^r \Phi_{br} \de^b_{\; p} + \partial^n \Phi_{np} + \ga_p A' \nonumber \\ &=& \ga_p A' \nonumber
\eea
for some $A$, $A'$.  Similarly, the third term in (\ref{expression}) is also a gauge transformation,
$$ \partial_b \Phi \ga_p^{\; \; b} = - \frac{1}{2}\partial^b \Phi \ga^b \ga_p + \ga_p B = - \partial_p \Phi+ \ga_p B', $$
for certain $B$ and $B'$, which yields also a gauge transformation of $\chi_p$.

So we  conclude that $2 \ga_5 \de_h \chi_p = 2 \partial_b \Phi_{pc} \ga^{cb} \ep $, i.e., 
\be
\de_h \chi_p = - \partial_b \Phi_{pc} \ga^{cb} \ga_5 \ep .\label{SUSY2ForChi}
\ee
Adding (\ref{SUSY1ForChi}) and (\ref{SUSY2ForChi}) together yields finally,
\be
\de \chi_p = \left( \partial_b P_{pc} - \partial_b \Phi_{pc}\ga_5 \right)  \ga^{cb} \ep .\label{SUSY3ForChi}
\ee

\subsection{Bosonic prepotential $Z^2_{mn} \equiv \Phi_{mn}$}
One has,
\begin{equation}
\delta h_{ij} = i \bar{\epsilon} \ga_i \psi_j + (i \leftrightarrow j). \label{VarMetric}
\end{equation}
Substituting the expression for $\psi_j$ yields
$$
\delta h_{ij} = i \bar{\epsilon} \ga_i \left(-   \ga_5  \partial^k \chi^p  \ep_{kjp}+  \ga_{js}  \ga_5 \partial_k \chi_p \ep^{ksp} \right)+ (i \leftrightarrow j).
$$
Using $\ga_i \ga_{js} = \delta_{ij} \ga_s - \delta_{is} \ga_j - \ep_{ijs} \ga_0 \ga_5$ and $\delta_{ij} \ep_{ksp} =  \de_{jk}\ep_{isp} + \de_{js} \ep_{ipk} + \de_{jp} \ep_{iks}$ one then gets,
$$
\delta h_{ij} =     i \ep_{jkp}   \partial^k (  \bar{\epsilon} \ga_i \ga_5 \chi^p  +   \bar{\epsilon} \ga^p \ga_5 \chi_i )+  \partial_j ( \bar{\epsilon} \ga^s \ga_5 \chi^p \ep_{isp}) + (i \leftrightarrow j).
$$
This implies, for the prepotential $\Phi_{mn}$ and the linearized diffeomorphism $u_m$,
\begin{eqnarray}
&& \delta \Phi_{mn} = i ( \bar{\epsilon} \ga_m \ga_5 \chi_n +  \bar{\epsilon} \ga_n \ga_5 \chi_m), \\
&& \de u_m = i    \bar{\epsilon} \ga^s \ga_5 \chi^p \ep_{msp}.
\end{eqnarray}

\subsection{Bosonic prepotential $Z^1_{mn} \equiv P_{mn}$}
The supersymmetry variation of the conjugate momentum $\pi^{ij}$ is most easily obtained from the supersymmetry variation of the metric and the equations of motion. 

One has, 
$$
\de( \partial_0 h_{ij}) = i \bar{\ep} \ga_i \partial_0\psi_j + (i \leftrightarrow j),
$$
and,
$$ 
\de (\partial_j h_{i0}) = i \bar{\ep} \ga_i \partial_j\psi_0 + i \bar{\ep} \ga_0 \partial_j\psi_i.
$$

Using the equations of motion (\ref{EoMRS}) for $\psi_j$ and the identity for $\ga_{i} \ga_{jm}$ just recalled above yields,
$$
\delta K_{ij} = \frac{i}{2} \bar{\ep} \ga_0 \partial_j \psi_i + \frac{i}{2} \ep_{jrs} \bar{\ep} \ga_i \ga_5 \partial^r \psi^s - \frac{i}{4} \de_{ij}  \ep^{mrs} \bar{\ep} \ga_m \ga_5 \partial_r \psi_s + (i \leftrightarrow j).
$$
This implies,
$$
\de \pi^{ij} = - \frac{i}{2} \bar{\ep} \ga_0 \partial^i \psi^j -  \frac{i}{2} \ep^{irs} \bar{\ep} \ga^j \ga_5 \partial_r \psi_s + \frac{i}{2} \de^{ij} \bar{\ep} \ga_0 \partial^k \psi_k+ (i \leftrightarrow j),
$$
for the conjugate momenta.

Now, we substitute in the variation of $\pi^{ij}$ the expression (\ref{SolForPsi}) for $\psi_m$ in terms of the spinor prepotential and expand the matrices occurring in the resulting equation in the basis $\{I, \ga_\mu, \ga_{\mu \nu}, \ga_{\mu \nu} \ga_5, \ga_\mu \ga_5, \ga_5 \}$ using the relevant identities.  One expects a priori terms involving $\ga_0 \ga_5$ or $\ga_k$. However, a direct computation shows that the terms containing $\ga_0 \ga_5 $ actually cancel, and that $\de \pi^{ij}$ reduces to
\bea
\de \pi^{ij} &=& - \frac{i}{2} \bar{\ep}  \ga^a \partial^i \partial_k \chi_p \ep^{j}_{\; \; sa} \ep^{ksp}  + (i \leftrightarrow j) \nonumber \\
&& + \frac{i}{2} \de^{ij} \bar{\ep}  \ga^a \partial_m \partial_k \chi_p \ep^{m}_{\; \; sa} \ep^{ksp}  + (i \leftrightarrow j)\nonumber \\
&&- \frac{i}{2} \bar{\ep}  \ga^j \partial_r \partial^k \chi_p \ep^{irs} \ep_{ksp}  + (i \leftrightarrow j)  \nonumber \\
&& + \frac{i}{2} \bar{\ep}  \ga_s \partial_r \partial_k \chi_p \ep^{irs} \ep^{jkp}  + (i \leftrightarrow j) .\nonumber
\eea
By expanding the products of $\ep^{bcd}$'s, one easily verifies that the sum of the first three terms is equal to the last term, so that,
$$
\de \pi^{ij} = i \bar{\ep}  \ga_s \partial_r \partial_k \chi_p \ep^{irs} \ep^{jkp}  + (i \leftrightarrow j),
$$
an expression which we can rewrite as
$$
\de \pi^{ij} = i \ep^{irs} \ep^{jkp} \partial_r \partial_k \left( \bar{\ep}  \ga_s \chi_p + \bar{\ep}  \ga_p \chi_s\right).
$$
The searched-for variation of the prepotential $P_{mn}$  then follows,
\be
\delta P_{mn} = i ( \bar{\epsilon} \ga_m \chi_n +  \bar{\epsilon}  \ga_n \chi_m).
\ee

\end{document}